\documentclass[epsfig,usegraphicx,useAMS,usenatbib]{mn2e}

\def\sqdeg{\,deg$^2$}
\def\etal{{et~al.~}}

\def\msol{\,{\rm M_{\odot}}}

\def\kpc{\,{\rm kpc}}

\def\mpc{\,{\rm Mpc}}

\def\kms{{\rm \,km \, s^{-1}}}
\def\Kms{$\kms$}
\def\mpc3h3{\,{\it h}^{-3}\, {\rm Mpc}^{3}}
\def\G3C{G$^3$Cv1}

\title[GAMA: In Search of MMAs]{Galaxy And Mass Assembly (GAMA): In Search of Milky-Way Magellanic Cloud Analogues}
\author[A.S.G. Robotham~\etal]{A.S.G.~Robotham$^{1,2}$\thanks{E-mail:asgr@st-and.ac.uk},
I.K.~Baldry$^3$,
J.~Bland-Hawthorn$^4$,
S.P.~Driver$^{1,2}$,
J.~Loveday$^5$,
\newauthor
P.~Norberg$^6$,
A.E.~Bauer$^7$,
K.~Bekki$^2$,
S.~Brough$^7$,
M.~Brown$^8$,
A.~Graham$^9$,
\newauthor
A.M.~Hopkins$^7$,
S.~Phillipps$^{10}$,
C.~Power$^2$,
A.~Sansom$^{11}$
L.~Staveley-Smith$^2$\\\\
$^1$SUPA\thanks{Scottish Universities Physics Alliance}, School of Physics \& Astronomy, University of St Andrews, North Haugh, St Andrews, KY16 9SS, UK\\
$^2$ICRAR\thanks{International Centre for Radio Astronomy Research}, The University of Western Australia, 35 Stirling Highway, Crawley, WA 6009, Australia\\
$^3$Astrophysics Research Institute, Liverpool John Moores University, Egerton Wharf, Birkenhead, CH41 1LD, UK\\
$^4$Sydney Institute for Astronomy, School of Physics, University of Sydney, NSW 2006, Australia\\
$^5$Astronomy Centre, University of Sussex, Falmer, Brighton BN1 9QH, UK\\
$^6$Institute for Computational Cosmology, Department of Physics, Durham University, South Road, Durham DH1 3LE, UK\\
$^7$Australian Astronomical Observatory, PO Box 296, Epping, NSW 1710, Australia\\
$^8$School of Physics, Monash University, Clayton, Victoria 3800, Australia\\
$^9$Centre for Astrophysics and Supercomputing, Swinburne University of Technology, Hawthorne, Victoria, Australia\\
$^{10}$HH Wills Physics Laboratory, University of Bristol, Tyndall Avenue, Bristol, BS8 1TL, UK\\
$^{11}$Jeremiah Horrocks Institute, University of Central Lancashire, Preston PR1 2HE, UK\\
\vspace*{-2em}
}

\begin{document}

\date{\vspace*{-4em}\noindent12/04/2011}

\pagerange{\pageref{firstpage}--\pageref{lastpage}} \pubyear{2011}

\maketitle

\label{firstpage}

\begin{abstract}

Analysing all Galaxy and Mass Assembly (GAMA) galaxies within a factor two ($\pm 0.3$ dex) of the stellar mass of the Milky Way (MW), there is a 11.9\% chance that one of these galaxies will have a close companion (within a projected separation of 70 kpc and radial separation of 400 km/s) that is at least as massive as the Large Magellanic Cloud (LMC). Two close companions at least as massive as the Small Magellanic Cloud (SMC) are rare at the 3.4\% level. Two full analogues to the MW-LMC-SMC system were found in GAMA (all galaxies late-type and star forming), suggesting such a combination of close together, late-type, star-forming galaxies is rare: only 0.4\% of MW mass galaxies (in the range where we could observe both the LMC and SMC) have such a system. In summary, the MW-LMC-SMC system is a 2.7$\sigma$ event (when recast into Gaussian statistics).

Using cross-correlation comparisons we find that there is a preference for SMC-LMC binary pair analogues to be located within 2 Mpc of a range of different luminosity groups. There is a particular preference is for such binaries to be located near LG luminosity systems. When these groups are subdivided into small magnitude gap and large magnitude gap subsets, the binaries prefer to be spatially associated with the small magnitude gap systems. These systems will be dynamically less evolved, but still offer the same amount of gravitational dark matter. This suggests that binaries such as the SMC-LMC might be transient systems, usually destroyed during vigorous merger events. Details of a particularly striking analogue to the MW-SMC-LMC and M31 complex are included.

\end{abstract}


\vspace{-20mm}

\section{Introduction}

The Local Group (LG), and more specifically the Milky-Way (MW) is the most thoroughly explored dark matter complex in the Universe \citep[e.g.][]{mate98,berg00,bens02,kara09,font11}. However, question marks remain over how typical the MW halo is in the context of the Universe and how unusual its galaxy occupation statistics are \citep[e.g.][]{boyl11,love11,toll11,weis11}. It is important that we fully understand how representative the MW halo is since, by virtue of proximity, it will always be the environment that will contain the faintest known galaxies and the broadest range of galaxy masses. It will also be the halo from which we can derive the most information about its formation history. Knowing which satellites of the MW halo are typical within similar-mass similar-redshift haloes will either severely tighten or relax the predictive requirements of N-body semi-analytic galaxy formation codes. Currently it is acknowledged that simulations struggle to predict the full distribution of MW satellite galaxies, these problems are particularly manifest for the brightest satellites: Large and Small Magellanic Clouds (LMC and SMC) \citep[e.g.][]{bens02,kopo09,okam10}.

We are set to learn a vast amount about the MW halo in the coming decades. In the near future GAIA \citep{wilk05} will measure space motions and properties for 2 billion stars in the LG which includes all known member galaxies. Amongst likely discoveries, we will learn about dynamical equilibrium, or lack of it, for the first time. Building up to these hugely detailed surveys it is important we discover where the MW halo fits into the bigger picture. Only then can we apply what we know about the MW to the $\Lambda$CDM (or some variant) model of the Universe. Combining near-field cosmology (LG scale) and far-field cosmology (redshift surveys) is key to completing the full picture of galaxy formation \citep{free02}.

This work puts the investigation of the MW halo into an observational cosmological context by using data from the Galaxy and Mass Assembly project (GAMA). GAMA is a multi-wavelength photometric and spectroscopic survey, and was designed to answer questions about how matter has assembled on a huge variety of scales: filaments, clusters, groups and galaxies. The first phase of the redshift survey was conducted on the AAT (known as GAMA-I) and these data are used in this work \citep{driv11}. In this work we use GAMA redshifts to search for close companions to MW mass galaxies. These systems will be Milky-Way Magellanic Cloud Analogues (MMAs from here). We use this sample to construct statistics on the rarity of SMC and LMC type (star forming late-type galaxies) close companions around $L^*$ late-type moderate star-formation rate spiral galaxies like our own MW.

In Section \ref{sec:Data} we discuss the data used in this work in detail. In Section \ref{sec:LGA} we present the statistics for finding MW-LMC and MW-LMC-SMC type systems, allowing us to quantify the apparent rarity of MW like systems. In Section \ref{sec:Cross} we investigate the environment that SMC-LMC type binaries are most commonly located in, and relate this to some of the defining characteristics of the LG.

Data for the LG was calculated using distance indicators without any $H_0$ dependence. As such it is appropriate (and consistent with the main body of LG literature) to convert GAMA data into true distance. To make the appropriate conversions we take the latest WMAP 7 cosmology: $\Omega_M=0.27$, $\Omega_\Lambda=0.73$ and $H_{0}=70\,{\rm km}\,{\rm s}^{-1}$ \citep{koma11}.

\vspace{-8mm}

\section{Data}
\label{sec:Data}

The Galaxy and Mass Assembly project (GAMA) is a major new
multi-wavelength spectroscopic galaxy survey \citep{driv11}.
The first 3 years of data obtained is referred to as GAMA-I, and is the data used for this work.
The GAMA-I data used here contain $\sim$86K redshifts to $r=19.4$
over $\sim$144~\sqdeg, with a survey design aimed at providing an
exceptionally uniform spatial completeness and high close pair completeness \citep{robo10a,bald10,driv11} . 

Extensive details of the GAMA survey characteristics are given in
\citet{driv11}, with the survey input catalogue described in
\citet{bald10} and the spectroscopic tiling algorithm in
\citet{robo10a}. Additional data used for this work
are stellar masses \citep{tayl11} and \G3C groups \citep{robo11}.

Table \ref{tab:LGinfo} shows the important MW, LMC and SMC values used for this work including distance and radial/tangential velocity separations to the MW. Conversions to the native GAMA $r$-band are made using the Lupton transforms.\footnote{\tt http://www.sdss.org/dr6/algorithms/sdssUBVRITransform.html}

\begin{table}
\begin{center}
\begin{tabular}{lrrrrrrr}
  \hline
  Name &  $M_V$ & $B-V$ & $M_r$ & $\mathcal{M}$ & $D$    & $V_R/V_T$  \\ 
  	    &               &		  &	         & $\msol$	   & kpc & km/s  \\
  \hline \hline
  MW  & -20.90$^a$ & 0.9$^b$ & -21.17 & $5 \times 10^{10}$$^c$     &  0  &  0  \\ 
  LMC         & -18.50$^a$ & 0.5$^d$ & -18.60  & $2.3\times10^{9}$$^e$ &  50$^f$  &  89/367$^f$  \\ 
  SMC         & -17.10$^a$ & 0.5$^d$ & -17.20  & $5.3 \times 10^{8}$$^e$ &  60$^f$ &  17/301$^f$   \\ 
   \hline \hline
\end{tabular}
\end{center}
\caption{\small Data used in this work. Data origin: \citet{berg00}$^a$, assumed M31 values for MW property$^b$, \citet{flyn06}$^c$, Nasa Extragalactic Database$^d$, \citet{jame11}$^e$, \citet{nich11}$^f$. All magnitudes use are intrinsic, i.e.\ $H_{0}=70\,{\rm km}\,{\rm s}^{-1}$. $M_{r}$ is derived using $M_{V}$ and the $B-V$ colour using the relevant Lupton SDSS photometric transform equation$^1$. }
\label{tab:LGinfo}
\end{table}

\vspace{-8mm}

\section{Finding MW Magellanic Cloud Analogues}
\label{sec:LGA}

There are a number of questions that could be stated with respect to investigating MMAs. For clarity this Section answers the following: `how common is it to observe 1 or 2 relatively large (stellar mass $> 1\times10^8 \msol$), star-forming satellites close to galaxies with stellar masses within 0.3 dex of the MW?' While this question lacks simplicity, it is at least possible to answer it in a meaningful, and reproducible, manner. This is a pertinent question to ask, since galaxy formation models have trouble replicating the presence of very bright satellites in close proximity to simulated galaxies like the MW \citep[e.g.][]{bens02}.

First we define `similar' stellar mass to mean within 0.3 dex of the MW mass \citep[$\mathcal{M}_{s-MW}=5\times10^{10} \msol$][]{flyn06}. Similarly we have to quantify `close'. A number of studies have been contacted that allow us to estimate the SMC/LMC distance and radial/tangential velocity components. Much endeavour has been invested in proper motion measurements by multiple teams \citep[e.g.]{kall06a,kall06b,cost09,viei10,cost11} which has allowed the more difficult to observe tangential velocity components to be estimated. These numbers have been further refined by detailed simulations citep[e.g.]{bekk08,nich11}. The SMC is $\sim60\kpc$ from the MW and travelling radially away at $\sim 17 \kms$ and $301\kms$ tangentially with respect to the MW: a net velocity of $302 \kms$ \citep{nich11}. The LMC is $\sim50\kpc$ from the MW and travelling away at $\sim 89 \kms$ radially and $367 \kms$ tangentially with respect to the MW: a net velocity of $378 \kms$ \citep{nich11}. To conservatively recover all systems where the galaxies are in such spatial--velocity proximity, we create a catalogue of pairs for this work where the projected separation is $r_{\rm sep-proj}<70\kpc$ and the radially velocity separation is $v_{\rm sep-rad}<400 \kms$. It should be stressed that whilst varying the precise definitions of `close' and `similar' does impact the total number of systems recovered and the fractions, it has minimal impact on the ratios between different population fractions.

A consequence of such a selection, and any similar, is that we do not truly distinguish between systems that have close pairs where all three galaxies show signs of independent 3-body interactions \citep[e.g. the M81-M82-NGC3077 group;][]{yun94} and systems like the MW-LMC-SMC that had a binary infall formation history \citep[e.g.][]{bekk08, nich11}. In data of GAMA quality there is too little phase space information (two dimensions of high accuracy spatial positions and one dimension of low accuracy velocities) to constrain the likely formation history of any system given the selection criteria stated above. Therefore, any fractions quoted should be considered as upper limits for finding systems that had a similar binary infall history to the MW-LMC-SMC but real limits for finding systems with exceptionally small phase separations regardless of the specific formation mechanism (e.g. both MW-LMC-SMC and M81-M82-NGC3077).

A second point to consider is how sensitive we are to the instantaneous flux of the SMC and LMC, both of which have had complicated star formation histories \citep[e.g.][]{harr04}. Since the complicated tidal interactions between the MW-LMC-SMC are known to trigger a large amount of the star formation \citep[e.g.][]{zari04}, it seems prudent to use the current luminosities and phase positions of all galaxies concerned in order to find analogues. This simplifies the search compared to looking at larger distances because such systems might have entirely different stellar mass content due to experiencing a more quiescent evolutionary history. In fact \citet{zari04} suggest that as much as 70\% of the stellar content of the SMC may have been formed due to interaction triggered star formation.

We use the $r<19.4$ GAMA-I survey data. Applying these selection limits to recover all systems that have similar pairwise properties to the MW and the Magellanic Clouds creates a catalogue containing 3,731 galaxy--galaxy pairs and 6,840 unique galaxies with no redshift limits applied. Obviously some galaxies will have more than one pair (the MW has two--- the SMC and LMC). We create complexes that contain all galaxy--galaxy associations and count this as a single `pair' system, i.e.\ the MW-LMC-SMC as a single `pair' system. Throughout we use the beta distribution to put robust estimates on the sampling statistics \citep{came11}, giving us firm limits on the 68 percentile probability range where relevant.

In the LMC depth sample ($0.01 < z_{\rm pair} < 0.089$ for $r<19.4$ mag) we find 1,642 galaxies that have $\mathcal{M}_{s}=\mathcal{M}_{s-MW}$ within a 0.3 dex range. Of these 286 galaxies are the dominant galaxy in the system (240 pairs, 39 triplets, 6 quartets, 1 quintet), where there are 340 minor galaxies and 626 galaxies in total. This suggests that 17.4\% of MW mass galaxies have at least one `close' companion. Of these paired systems 56/286 have late-type dominant pair galaxies (19.6\%). Of all the systems that have a late-type brightest pair galaxy (BPG), 34 of the minor pair galaxies are late-type. Since complexes can contain more than one pair, this translates as 31/56 late-type BPGs have {\it at least} one late-type minor companion (55.4\%) and 30/56 that have 100\% late-type companions (53.6\%).

Of these 30 systems that are 100\% late-type, all have some observable H$\alpha$ emission in the larger galaxy (100\%) and 32/34 minor companions have some amount of observable H$\alpha$ emission (94.1\%). Comparing to the larger sample of MW mass selected galaxies this means that 30/1,642 (1.8\%, beta range 1.5\%--2.2\%) of all MW mass galaxies are late-type and in a pair, where all of the minor galaxies are late-type and all galaxies are star forming. Thus, approximate analogues of the MW system (where we just require all close companions to be Magellanic Cloud like) are rare at the less than 2\% level. In this cascade of fractions, the most unusual characteristic is to find a late-type MW mass galaxy in a pair at all, followed by the dominant galaxy being late-type given that it is in a pair. Once these criteria have been met the chance of finding a late-type companion, and star formation in both galaxies, is remarkably high (over 50\%). This means the discussion of the uniqueness of the MW system is largely driven by how rare galaxy pairs are at low redshift.

Three triple systems are present in the final selection. The effective stellar mass limit is less well defined than the $r$-band limit (which has to assume an approximate k-correction), but based on when the number counts begin to turn over for galaxies with stellar mass less than the SMC the survey is complete out to $z \sim 0.055$. The total observable volume, using the standard cosmology of this paper, is $1.8 \times 10^{5}$Mpc$^3$. Two of the three triple systems fall within this redshift range. Since 414 MW  stellar mass $\pm 0.3$ dex galaxies are within this redshift limit, full analogues of the MW-LMC-SMC system are rare at the 0.4\% level (0.3\% to 1.1\%). In terms of space density, we find $1.1 \times 10^{-5}$Mpc$^{-3}$ full analogues in GAMA (in a volume of $1.8 \times 10^{5}$Mpc$^3$). It is 6 times more likely that a MW mass galaxy with two SMC mass companions will have early-type morphology. Casting this figure in terms of Gaussian statistics, full analogues of the MW halo are rare at the 2.7$\sigma$ level when searching around $L*$ galaxies.

Of the two systems, only one has a minor companion close to the stellar mass of the SMC (the other has two LMC mass companions). This system is the nearest analogue to the MW system found in the GAMA database, possessing a dominant star forming spiral galaxy with $\mathcal{M}_{s} = 3.1 \times 10^{10} \msol$ ($\mathcal{M}_{s-MW} = 5 \times 10^{10} \msol$), a more massive late-type companion with $\mathcal{M}_{s} = 6.1 \times 10^{9} \msol$ ($\mathcal{M}_{s-LMC} = 2.3 \times 10^{9} \msol$) and a less massive late-type companion with $\mathcal{M}_{s} = 6.1 \times 10^{8} \msol$ ($\mathcal{M}_{s-SMC} = 5.3 \times 10^{8} \msol$). Both of the companions are more massive than their Magellanic Cloud equivalents, but the smaller one is very close to the mass of the SMC.

\begin{figure*}
\centerline{\mbox{\includegraphics[width=7in]{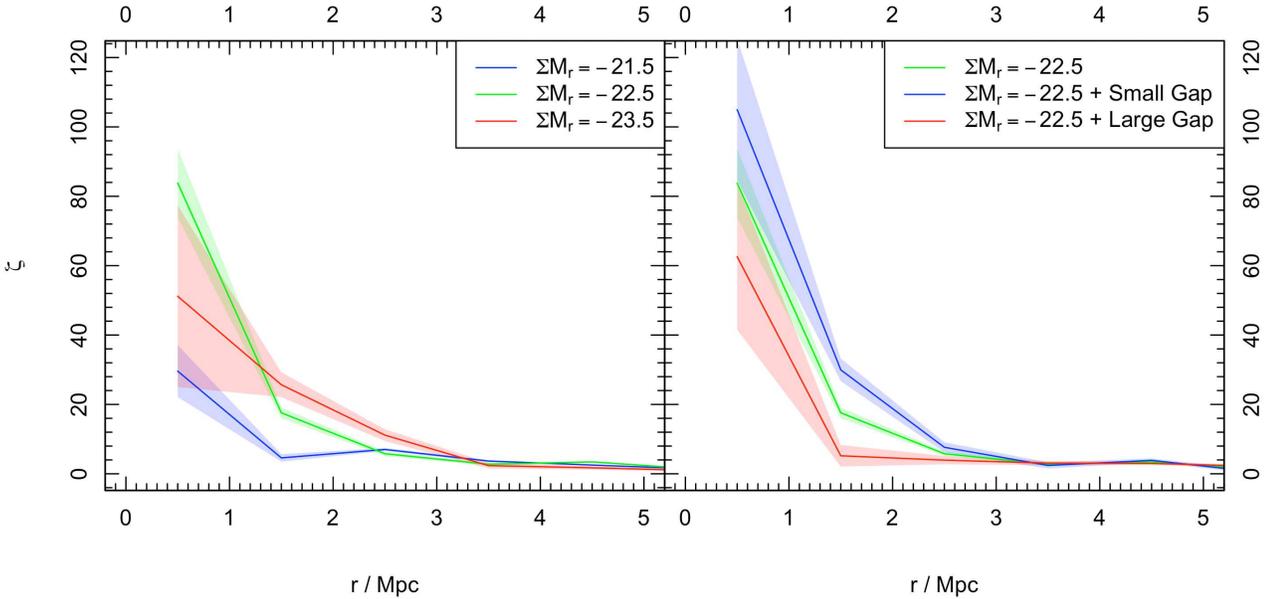}}}
\caption{\small Left panel shows the cross-correlation between close binaries that are analogues to the SMC / LMC pair (Magellanic Cloud Analogues: MCAs) and \G3C groups of different extrapolated flux content within 0.5 mag of the value stated in the legend. The total $r$-band magnitude of the Local Group is $M_{r-LG}= \sim -22.2$, and the total magnitude of the MW halo is $M_{r-MW}= \sim -21.3$. The right panel splits the most highly clustered MCA-group subset ($\Sigma M_r = -22.5$) into two further subsets: one with $M_{r-2^{nd}}-M_{r-1^{st}}$ less than the median magnitude gap value of 0.7 for the groups selects, and another with $M_{r-2^{nd}}-M_{r-1^{st}}$ larger than the median. In both plots the y-axis is the standard cross-correlation excess, and is the relative excess compared to the random volume.}
\label{fig:crosscor}
\end{figure*}

While the numbers found in this work suggest the MW-Magellanic Cloud system is cosmologically rare, we at least know that such a combination of galaxy diagnostics is not entirely unique. It is interesting to note that this system does not find the two companions to be in a close binary (as the LMC and SMC appear to be). This is similar to the findings of \citet{jame11} who, using an H$\alpha$ limited imaging survey of 143 spiral galaxies, did not find a single MW-Magellanic Cloud analogue that had two companions in a close binary formation. This suggests that the binary nature of the Magellanic clouds might be their defining unique feature.

To compare to the analysis of the Millenium II simulation by \citet{boyl11}, we select the 1,642 galaxies that are within 0.3 dex of $\mathcal{M}_{s-MW}$ as discussed above. We now apply a mass selection on the pairs. If we state that any minor pair galaxy must be the mass of the LMC or larger ($\mathcal{M}_{s-LMC} = 2.3 \times 10^{9} \msol$) then we find 196 galaxies that have a sub-halo occupied by a close companion that is at least that massive. This means that given a halo that is occupied by a MW mass galaxy, there is a 11.9\% (11.2\%--12.8\%) chance that a galaxy at least as massive as the LMC occupies a sub-halo.

\citet{boyl11} find that in haloes that have a MW mass galaxy there will be a sub-halo containing a galaxy at least as massive as the LMC 3--8\% of the time if the halo is $\sim 1 \times 10^{12}\msol$ and 20-27\% of the time if the halo is $\sim 2.5 \times 10^{12}\msol$. The median \G3C halo mass we find for MW mass galaxies is $\sim 2 \times 10^{12} \msol$ (using the functional A scaling in Robotham et al. 2011). This number is consistent with the halo mass of the MW given by \citet{li08}. Our probability range of LMC or more massive sub-haloes (11.2\%--12.8\%) is also between the ranges stated by \citet{boyl11}, suggesting it is broadly consistent with their results.

To calculate similar statistics for high order systems we apply the $0.01<z<0.055$ SMC stellar mass depth limit (leaving 414 galaxies), and remove galaxies with stellar mass less than the SMC. We find 14/414 MW mass galaxies that have 3 or more galaxies in the pair system, i.e.\ 3.4\% (2.7\%--4.5\%) of systems where we expect to be able to observe both the LMC and SMC have at least 2 galaxies with stellar mass greater than the SMC. The constraint is less tight due to poorer number statistics, a consequence of the smaller parent population within the SMC observable redshift limit. Purely framing the discussion in terms of how likely it is for baryons to occupy sub-haloes, these results imply that, assuming the dominant galaxy in the halo is similar in mass to the MW, it is 3.5 times less likely to have two sub-haloes with at least SMC stellar mass compared to 1 sub-halo with at least LMC stellar mass.

\vspace{-8mm}

\section{Where do Magellanic Clouds Analogues Live?}

\label{sec:Cross}

An actively discussed mechanism for explaining the presence of the Magellanic Clouds in the MW halo is binary infall \citep[see discussion in][]{bekk08, kall09,yang10,nich11}. This model assumes the SMC and LMC were a binary pair that entered the MW-halo / LG-complex simultaneously. Observationally we can determine the viability of such a mechanism by searching first for close binary analogues to the Magellanic Clouds, and then determining the cross correlation these systems have with various luminosity groups taken from the GAMA Galaxy Group Catalogue \citep[\G3C,][]{robo11}.

To select close binary Magellanic Cloud Analogues (MCAs) we search for all galaxies that have a close companion within a projected separation of 100 kpc and velocity separation of 100 km/s, where both galaxies are between $-19 < M_r < -17$. This selection conservatively selects all binary pairs with major characteristics in common with the SMC and LMC. 46 such binary MCA systems are found, where 1,929 galaxies fall inside the redshift and magnitude selection limits. This implies $4.8\%$ (4.3\%--5.3\%) of galaxies in the specified magnitude range are in such close binary systems.

Taking the \G3C catalogue we use the cross-correlation approach of \cite{crof99} to determine how spatially associated the MCAs are with different luminosity groups (within 0.5 mag of the stated values), where we use the extrapolated $r$-band flux content given in the \G3C catalogue. Errors are estimated through jack-knife resampling (measuring the variance in the cross correlation signal when binary pairs are excluded) and random volume cones are generated through uniform pointing (within the redshift range explored the sample is complete). The left-hand panel of Figure \ref{fig:crosscor} shows how closely associated the MCAs are to different types of groups. The medium galaxy flux plotted on the left panel (green line) includes groups with the same flux content as the entire LG. The red line shows more massive systems than the LG, the blue line includes groups with the same total flux as just the MW halo. It is clear that MCAs are more likely to be found in proximity to LG mass complexes, suggesting that the presence of M31 near to the MW halo has enhanced the probability of observing a binary system like the SMC-LMC. A caveat to this result is the cross-correlations are affected by the subset comparisons chosen, and given the known interplay between luminosity and interactions the systems detected at large separations are likely to be in a different evolutionary state to those at close separations.

To further investigate the effect seen for $\Sigma M_r = -22.5$ groups this cross-correlation was subdivided into two sub-sets: one where the magnitude gap of the brightest two galaxies ($M_{r-2^{nd}}-M_{r-1^{st}}$) is larger than the median value of 0.7 (so the central galaxy dominates) and another where the magnitude gap is smaller than the median (so the central galaxy does not dominate). The right-hand panel of Figure \ref{fig:crosscor} shows the cross-correlations for these two subdivisions, with the original full sample plotted also. There is a marginally significant preference for MCAs to be more spatially associated with groups with small magnitude gaps. These systems should be dynamically less evolved since the central galaxy is not as dominant within the group, and in fact is one of the major characteristics of the LG: the MW and M31 are similarly massive galaxies that would be found in the `small gap' sample. This data lends evidence that the MW halo should not be considered in isolation when determining the occupation probability of the Magellanic Clouds. The potential role of M31 on the presence of the LMC and SMC near the MW has been considered in recent simulation work \citep[e.g.][]{kall09,yang10}, suggesting that its presence might be significant rather than coincidental.

To quantify this effect differently we move from considering the cross correlation of MCAs with groups, to directly searching for $L^*$ pairs in close proximity. We create the $L^*$ pair sample by searching for all galaxies between $-21.9 < M_r < -20.9$ (the MW and M31 fall well inside this selection), within $r_{\rm proj} < 1,000 \kpc$ and $v_{\rm sep} < 500 \kms$. Of the 302 galaxies that fall within the magnitude selection 96 (32\%) have a close $L^*$ companion (i.e.\ there are 48 pair systems). For each pair system we calculate the effective $r$-band centre-of-light in RA, Dec and redshift and define this as the centre of the pair system. We now search for all MCAs that are within a $r_{\rm proj} < 1,000 \kpc$ and $v_{\rm sep} < 500 \kms$ separation to $L^*$ pair systems. 11/47 MCAs are found in close proximity to $L^*$ pair systems, while 27/47 MCAs are found within the same spatial separation of {\it any} $L^*$ galaxy. At a maximum the effective comoving volume searched over for all $L^*$ systems is $1.3 \times 10^4$ Mpc$^3$, which is $\sim$7\% of the available GAMA volume within these redshift limits. For the $L^*$ pairs the maximum volume searched is $2.1 \times 10^3$ Mpc$^3$, which is $\sim$1\% of the available GAMA volume. Consequently we find that 57\% of MCAs are found within 7\% of the volume when searching around $L^*$ systems, and 23\% of MCAs are found within 1\% of the volume when searching around the centres of $L^*$ pairs. There is a clear tendency for MCAs to be associated with $L^*$ systems in general (rather than just random distributed throughout the Universe), and an even stronger association is seen between MCAs and binary pairs of $L^*$ galaxies, like the MW and M31.
 
\begin{table}
\begin{center}
\begin{tabular}{lrrrrrrr}
  \hline
  GAMA ID	& RA (J2000) 	& Dec (J2000) 	& Redshift 	& $m_r$ \\
  \hline \hline
  MMA & & & & \\
  \hline
 202627		& 08:42:28.28	& -00:16:17.8	& 0.05130		& 15.25 \\
 202636		& 08:42:28.13	& -00:17:00.7	& 0.05134		& 17.25 \\
 202691		& 08:42:30.64	& -00:16:22.3	& 0.05119		& 18.47 \\
 \hline
 Close Spiral	&			&			&			&	\\
 \hline
 202673		& 08:42:36.66	& -00:13:51.5	& 0.04991		& 15.31 \\
   \hline
\end{tabular}
\end{center}
\caption{\small Basic information for a particularly striking example of a Milky-Way Magellanic Cloud Analogue (GAMA-MMA1) with a near-by bright spiral companion. This mimics a lot of the most recognisable characteristics of the MW-SMC-LMC and M31 complex.}
\label{tab:MMA}
\end{table}

\begin{figure}
\centerline{\mbox{\includegraphics[width=3.4in]{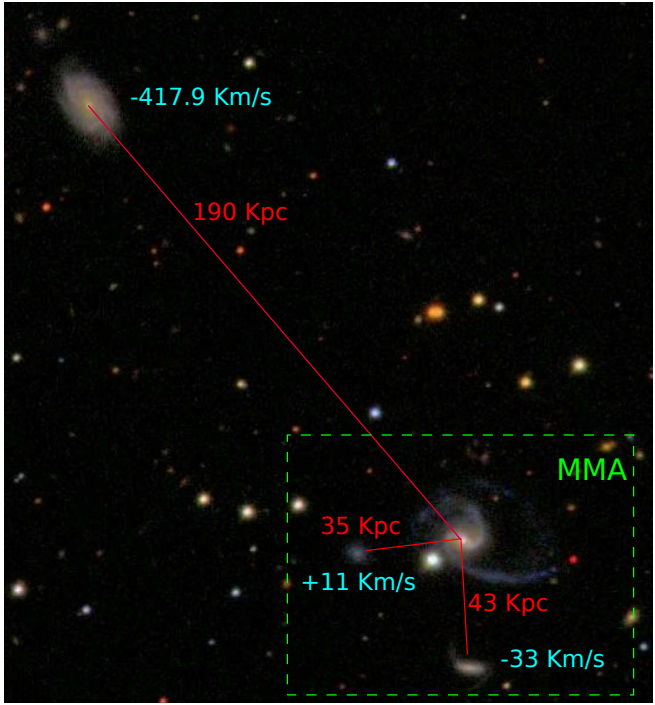}}}
\caption{\small SDSS image of the best LG analogue found in GAMA. The spiral galaxy in the bottom right with the two indicated brighter companions is the MMA (GAMA-MMA1), where all three galaxies are late-type and star-forming, making this system similar to the MW-Magellanic Cloud system in the most fundamental ways. An associated nearby spiral galaxy is also shown.}
\label{fig:BestLGA}
\end{figure}

The best MMA (constituting a spiral brightest-galaxy and star forming companions, known as GAMA-MMA1) also has a near-by companion spiral galaxy. A multi-colour image based on SDSS photometric data is shown in Figure \ref{fig:BestLGA}. Table \ref{tab:MMA} contains key information on the system. The companion spiral is 190 Kpc away in projection (MW and M31 are separated by 800 Kpc in real space) and  -400 \Kms separated in velocity (M31 is moving at $-122$\Kms radially and $\sim 100$\Kms tangentially with respect to the MW). These two spirals have $M_{r} = -21.43 / -21.31$, which is very similar to MW / M31 ($M_{r} = -21.47 / -21.17$). IFU and Parkes data have recently been obtained for this newly identified analogue to the LG system, and we are actively seeking followup on a sample of the MMAs discussed in this work in order to better quantify the occupation statistics in halos similar to our own.

\vspace{-8mm}

\section{Conclusions}

The major findings of this work are summarised below:

\begin{itemize}

\item Analysing all galaxies within 0.3 dex of the stellar mass of the MW, there is a 11.9\% (11.2\%--12.8\%) chance that it will have a close companion (within a projected separation of 70 kpc and radial separation of 400 km/s) that is at least as massive as the LMC. This is consistent with analyses by \citet{boyl11} of the Millenium II simulation and by \citet{jame11} of H$\alpha$ imaging around luminous spiral galaxies.

\item Limiting the sample to those galaxies where the SMC would be observable we find 3.4\% (2.7\%--4.5\%) of galaxies have two companions at least as massive as the SMC.

\item Only two full analogues to the MW-LMC-SMC system were found in GAMA, suggesting such a combination of late-type, close star-forming galaxies is quite rare: in GAMA only 0.4\% (0.3\%--1.1\%) of MW mass galaxies have such a system (a 2.7$\sigma$ event). In terms of space density, we find $1.1 \times 10^{-5}$Mpc$^{-3}$ full analogues in GAMA (in a volume of $1.8 \times 10^{5}$Mpc$^3$). The best example found shares many qualitative characteristics with the MW system. The brightest pair galaxy has spiral features, as does the bigger minor companion. The minor companions are $\sim$40 kpc in projected separation, so not in a close binary formation like the SMC and LMC.

\item Selecting systems that are close binaries like the SMC-LMC pair (MCAs), we find that they are preferentially located in close proximity (or within) systems that have a similar total flux to the LG ($\Sigma M_r = -22.5 \pm 0.5$ mag).

\item Subdividing the preferential group type into those with large and small magnitude gaps, we find MCAs are more spatially associated with groups that have a small magnitude gap. This suggests that a quiet recent merger history improves the likelihood of the Magellanic Clouds being visible in the LG. The best MMA analogue found in GAMA also has a close $L^*$ spiral companion galaxy.

\end{itemize}

\vspace{-8mm}

\section*{Acknowledgments}

ASGR acknowledges STFC and SUPA funding that was used to do this work. PN acknowledges a Royal Society URF and an ERC StG grant (DEGAS-259586). Thanks go to James Ivory for helpful comments at various stages of this work. Thanks also to the anonymous referee for their helpful comments. GAMA is a joint European-Australasian project based around a spectroscopic campaign using the Anglo-Australian Telescope. The GAMA input catalogue is based on data taken from the Sloan Digital Sky Survey and the UKIRT Infrared Deep Sky Survey. Complementary imaging of the GAMA regions is being obtained by a number of independent survey programs including GALEX MIS, VST KIDS, VISTA VIKING, WISE, Herschel-ATLAS, GMRT and ASKAP providing UV to radio coverage. GAMA is funded by the STFC (UK), the ARC (Australia), the AAO, and the participating institutions. The GAMA website is {\tt http://www.gama-survey.org/}.

\vspace{-8mm}

\bibliographystyle{mn2e}
\setlength{\bibhang}{2.0em}
\setlength\labelwidth{0.0em}
\bibliography{mmav1}

\label{lastpage}

\end{document}